\documentstyle[aps,prl,twocolumn]{revtex}

\newcommand{\mod}{{\rm \; mod\; }}

\begin{document}

\draft
\title{Commensurability, excitation gap and topology \\
in quantum many-particle systems on a periodic lattice}

\author{Masaki Oshikawa}

\address{
Department of Physics, Tokyo Institute of Technology,
Oh-okayama, Meguro-ku, Tokyo 152-8551, Japan
}

\date{August 9, 1999}

\maketitle

\begin{abstract}
Combined with Laughlin's argument on the quantized Hall conductivity,
Lieb-Schultz-Mattis argument is extended to quantum many-particle
systems (including quantum spin systems) with a conserved particle number,
on a periodic lattice in arbitrary dimensions.
Regardless of dimensionality, interaction strength and
particle statistics (bose/fermi),
a finite excitation gap is possible only when the particle
number per unit cell of the groundstate is an integer.
\end{abstract}


%\pacs{\bf Please do not distribute}

Strongly interacting quantum many-particle systems
is one of the central topics of theoretical physics.
The Renormalization Group (RG) is an important concept in studying
such problems~\cite{WilsonKogut}.
According to the RG picture,
low-energy, long-distance behavior of a quantum many-particle
system is governed by RG fixed points.
A quantum critical system, which has gapless excitation spectrum,
is renormalized into a critical RG fixed point.
However, in general, a critical RG fixed point allows some relevant
perturbations.
When the perturbations are added, the system is driven away from
the critical fixed point. Such system generally
has a finite excitation gap.
Then it is expected that a gapless quantum critical system is unstable and
only achieved by an appropriate fine-tuning of the Hamiltonian,
which makes the relevant perturbations vanish.

In reality, there are rather many quantum critical system with
gapless excitation spectrum, obtained without any apparent fine-tuning.
Thus, one may naturally ask a question: is there some mechanism
which protects a gapless critical system?
There is one well-known mechanism of such kind:
a gapless Nambu-Goldstone mode~\cite{NG} exists when the system has a
{\em spontaneously broken} continuous symmetry.
In fact, it describes variety of gapless excitations in 
quantum many-body systems, like spin waves and phonons, etc.
However, it does not exhaust all the (stable) quantum critical systems.
On the other hand, few universal mechanism other than
the Nambu-Goldstone theorem are known~\cite{Affleck:LesHouches}.

In this letter, we argue that an incommensurability is another
universal mechanism which protects the gapless excitation spectrum
in quantum many-particle systems.
We will show the following statement:
\begin{quote}
\em
In a quantum many-particle system defined on a periodic
lattice, with an exactly conserved particle number,
a finite excitation gap is possible only if
the particle number per unit cell
of the ground state is an integer.
\end{quote}

This condition to have a finite gap
may be called as the ``commensurability condition''.
When the particle number per unit cell~\cite{unitcell} of
the lattice is $\nu = p/q$ where $p$ and $q$ are coprimes,
a gapful groundstate must spontaneously break the translation symmetry,
so that the unit cell of the groundstate is enlarged by factor of $q$.
In the case of quantum spin system with the conserved total magnetization
$\sum_j S^z_j$, mapping of the spin system to an interacting
boson system gives $S-m$ particles per spin where $S$ is the spin
quantum number, $m$ is the average magnetization per spin.

An incommensurate filling corresponds to the limit of large $q$,
giving large ground state degeneracy
{\em if there is a finite excitation gap.}
This usually implies the spectrum is actually gapless. 
That the incommensurate filling gives a gapless spectrum
is empirically recognized more or less.
In fact, all trivial ground states (e.g. completely dimerized state)
with an excitation gap,
as well as the less trivial
Valence-Bond-Solid states~\cite{VBS}, 
satisfy the commensurability condition.
However, it is not trivial whether the incommensurability
generally guarantees gapless excitations in the presence of
interaction and quantum fluctuation.

If a gapful phase is adiabatically
connected to those trivial states within the Hamiltonians
conserving the particle number and periodicity,
it must also satisfy the same commensurability condition.
This is because an infinitesimal modification of the Hamiltonian
does not mix states with different particle number;
thus the particle number per unit cell in the groundstate
is unchanged during the adiabatic change, if the gap always remains open.
This explains the stability of the particle density $\nu$
in such cases (e.g. in the ``strong coupling'' approach to the
magnetization plateau~\cite{Hida}).
However, this argument does not exclude the possibility
of a gapful phase,
which is not adiabatically connected to any trivial state,
not obeying the commensurability condition.

In this letter, based on a topological argument,
we will show that the commensurability condition is
a robust non-perturbative constraint.
We consider general quantum many particle systems
on a lattice with a periodic structure with
any strength of interaction, in $D$-dimensions.
For simplicity, here we assume that there is a single species of
particles.
We assume that the number of particles is conserved (i.e.
commutes with the Hamiltonian.)
Let us call the direction of an arbitrary chosen primitive
vector $\vec{a}$ of the lattice as $x$-direction.
We impose the periodic boundary condition in $x$-direction
with length $L$ measured in unit of $|\vec{a}|$.
Defining the translation operator $T_x$ which
translates the system by $\vec{a}$,
the periodic boundary condition is represented as ${T_x}^L =1$,
and the Hamiltonian invariant under the translation $T_x$.
The ``cross section'' of the lattice is defined so that
the whole lattice is spanned (without overlap)
by translation of the cross section by $T_x$.
We denote the number of unit cells contained in the cross section
by $C$; the total volume (ie. number of unit cells)
of the system is given by $C L$.

In one dimension (including ladders etc.),
the proposed statement was already shown in Refs.~\cite{OYA,YOA}
by generalizing the Lieb-Schultz-Mattis (LSM) argument~\cite{LSM}.
Therefore the remaining problem is to understand higher dimensions.
Applying the LSM argument to the higher dimensions $D > 1$
meets a difficulty.
The energy of the variational state is bounded only by $O(C/L)$,
which is generally not small in the thermodynamic limit;
in an isotropic (in size) system, $C \sim L^{D-1}$.

Affleck~\cite{ALSM} discussed some application of
the LSM argument to $D>1$.
While in Ref.~\cite{ALSM} only spin systems at zero magnetization
were considered, it is straightforward to extend
the discussion in Ref.~\cite{ALSM}
to quantum many-particle systems with general
particle density (quantum spin systems with general magnetization),
as was done in one dimension~\cite{OYA,YOA}.
Unfortunately, the strong anisotropy limit $C/L \to 0$
is necessary to apply the LSM argument as in Ref.~\cite{ALSM}.
He argued it plausible that the conclusion is still valid
for an isotropic system where $C \sim L^{D-1}$.
However, the strong anisotropy limit makes the system
essentially one-dimensional.
Thus one might suspect that the LSM argument does not give
a useful information on higher-dimensional system which is
isotropic in size.
Below we will argue that the same conclusion holds, without
relying on the strong anisotropy limit.

As in Ref.~\cite{ALSM}, 
we impose the periodic boundary condition for $x$-direction,
and require $C$ to be mutually prime with $q$
but not the anisotropy condition $C \ll L$.
The boundary conditions for other than $x$-direction can be either
open or periodic if they are uniform in $x$-direction.
The particles may or may not have a real electric charge.
Here we introduce a fictitious charge $e$ for each particle,
which couples to an externally given (fictitious) electromagnetic field.

Because of the periodic boundary condition in the $x$-direction,
the system may be regarded as a ring.
Following Laughlin's discussion~\cite{Laughlin}
of the Quantum Hall Effect (QHE),
we consider a magnetic flux $\Phi$ piercing through the ring.
In a simplest gauge choice.
the magnetic flux can be represented by
the uniform vector potential $A_x = \Phi/L$ in the $x$-direction,
Now let us adiabatically increase the
magnetic flux $\Phi$ by a unit flux quantum $\Phi_0 = hc/e$,
when the system is in the groundstate $| \Psi_0 \rangle$ at $\Phi=0$.
The groundstate $|\Psi_0 \rangle$ is chosen
(when the groundstates are degenerate) so that it is also an eigenstate
of $T_x$.
This is always possible, at least in a finite size system, because
we assumed periodic lattice structure and periodic boundary condition
in $x$-direction; the Hamiltonian commutes with $T_x = e^{i P_x}$.
Here $P_x$ is the $x$-component of the total (crystal) momentum.
The groundstate is thus also an eigenstate of the momentum with an
eigenvalue $P_x^0$:
\begin{equation}
	P_x | \Psi_0 \rangle = P_x^0 | \Psi_0 \rangle,
\label{eq:pxgs}
\end{equation}

During the adiabatic process, the Hamiltonian is only modified
by the uniform vector potential $A_x = \Phi / L$
in the above gauge choice.
Then the Hamiltonian always commutes with $T_x$.
When the magnetic flux reaches the unit flux quantum,
the original groundstate evolves into some state
$| \Psi'_0 \rangle $.
The Hamiltonian $H(\Phi)$ generally depends on the flux $\Phi$
through the vector potential, reflecting the Aharanov-Bohm (AB) effect.
However, when the AB flux $\Phi$ is equal to the unit flux quantum,
there is no AB effect and the energy spectrum is identical
to the zero flux case.
In fact, the vector potential is eliminated by the
large gauge transformation
\begin{equation}
U = \exp{[ \frac{2 \pi i}{L} \sum_{\vec{r}} x n_{\vec{r}}]},
\end{equation}
where $n_{\vec{r}}$ is the particle number operator
at site $\vec{r}$, and $x$ is the $x$-coordinate of $\vec{r}$.
Namely, the Hamiltonian with the unit flux quantum
goes back to the original one by the large gauge
transformation as $U H(\Phi_0) U^{-1} = H(0)$.
By the same transformation, the adiabatic evolution of the
groundstate $| \Psi'_0 \rangle$ is mapped to
$U | \Psi'_0 \rangle$.
Thus, after the adiabatic procedure and the large gauge transformation,
we get back to the original Hamiltonian but the groundstate
$| \Psi_0 \rangle$ is changed to $U | \Psi'_0 \rangle$.

On the other hand, 
in the presence of a finite excitation gap,
the ground state $| \Psi_0 \rangle$
can only be transformed into itself
or, possibly, into another one of degenerate groundstates
during the adiabatic
process~\cite{Laughlin,TaoWu}.
As already explained,
the reason why the LSM argument has been applied only to one dimension
(or to the strong anisotropic limit) is that,
the energy expectation value of the variational state 
is bounded only by $O(C/L)$, which is generally not small.
Thus one can not say $U | \Psi_0 \rangle$ is always
a low-energy state in $D >1$ dimensions.
However, applying the adiabatic argument, we are able to claim
that, {\em if the system has a finite excitation gap},
the outcome of the adiabatic evolution $U | \Psi_0' \rangle$
should be one of the groundstates~\cite{note}.

In the case of QHE,
an implicit assumption~\cite{Laughlin}
of uniqueness of the groundstate led to an integer quantization
of Hall conductivity.
However, as pointed out by Tao and Wu~\cite{TaoWu},
it is possible that the ground states are degenerate,
and that is what needed in fractional QHE.
Therefore we have to check whether $U | \Psi'_0 \rangle$
is identical to $| \Psi_0 \rangle$ or not.

Here let us recall that, $U | \Psi_0 \rangle$,
which is similar to $U | \Psi_0' \rangle$, is
nothing but the variational state constructed
in the LSM argument and its
generalizations~\cite{LSM,AL,ALSM,OYA}.
Now we see a rather close relation between LSM and
Laughlin's arguments.
While our state $U |\Psi_0' \rangle$ is not identical to
the LSM state $U |\Psi_0 \rangle$,
we can still invoke the LSM orthogonality argument
used for $U | \Psi_0 \rangle$.

We have chosen the original groundstate as an eigenstate
of $P_x$ as in eq.~(\ref{eq:pxgs}).
Since the Hamiltonian always commutes with $P_x$ during the
adiabatic process, the eigenvalue of $P_x$ is unchanged.
This can be seen easily from perturbation theory
on an infinitesimal increase of the AB flux; the
infinitesimal modification of the Hamiltonian commutes
with $P_x$ and it does not mix states with different eigenvalues of $P_x$.
Thus, $| \Psi'_0 \rangle$ belongs to the same eigenvalue $P_x^0$
as in eq.~(\ref{eq:pxgs}).
Now, after the gauge transformation, $| \Psi'_0 \rangle$
is transformed to $U | \Psi'_0 \rangle$, which may belong
to a different eigenvalue.
By using the identity
\begin{equation}
U^{-1} T_x U = T_x \exp{[ 2 \pi i \sum_{\vec{r}} \frac{n_{\vec{r}}}{L} ]}
\end{equation}
we see that $U | \Psi'_0 \rangle$ is an eigenstate of $P_x$
with $P_x = P_x^0 + 2 \pi \nu C$.
Thus, if choose $C$ to be mutually prime with $q$ ($\nu = p/q$), 
$U | \Psi'_0 \rangle$ is orthogonal to $| \Psi'_0 \rangle$
and $| \Psi_0 \rangle$, because it belongs to a different eigenvalue
of $P_x$.
By repeating the same procedure several times,
we can conclude that there are at least
$q$ degenerate ground states.
Thus we have shown that the similar conclusion to the one-dimensional
case~\cite{OYA,YOA} holds in any dimensions,
without relying on the strong anisotropy limit.
The optimistic view taken in Ref.~\cite{ALSM} is actually justified,
as far as the anisotropy condition is concerned.
It is valid for arbitrary strong interaction, and is even
independent of the particle statistics (bose/fermi).
The ground state degeneracy is a robust, non-perturbative property
related to the topology of the gauge field
and the symmetries of the system.

An unsatisfactory point still remaining is that
we have to take $C$ to be mutually prime with $q$.
If $C$ is an integral multiple of $q$,
nothing can be said on the degeneracy, and a unique ground state
with an excitation gap is possible in principle.
This is also related to the fact that the present argument
is not yet mathematically rigorous for the thermodynamic limit.
(Compare to Ref.~\cite{AL}.)
However, if we assume that
the groundstate degeneracy does depend on whether
$C$ is an integral multiple of $q$ or not in a large enough system,
it suggests some long-range structure of period $q$.
Then it is naturally expected~\cite{ALSM}
that the ground states have
the $q$-fold degeneracy anyway, for a sufficiently large system.
In addition, the present argument can be applied to 
many different boundary conditions,
because there are various possible choices of the primitive vector $\vec{a}$
and the corresponding cross section.
The degeneracy looks less artificial in the light of this fact.

The ground states in a finite system would be actually split
by exponentially small energy due to tunneling effects.
In a finite size system, the $q$ (near-)groundstates $|\Psi_n \rangle$
($n=0,1,\ldots, q-1$) are eigenstates of the momentum $P_x$:
$P_x | \Psi_n \rangle = (2 \pi n / q) | \Psi_n \rangle$.
However, in the thermodynamic limit, the physical groundstates
are given by $| \tilde{\Psi}_j \rangle$'s, which are defined as
$| \tilde{\Psi}_j \rangle = \sum_n e^{i P_x j} | \Psi_n \rangle$.
These physical ground states are connected by the translation operator:
\begin{equation}
T_x | \tilde{\Psi}_j \rangle = | \tilde{\Psi}_{(j+1) \mod n} \rangle .
\end{equation}
Thus the translation symmetry (to $x$-direction) is spontaneously broken,
and the periodicity of the ground state is an integral multiple of $q$.
This concludes the derivation of the proposed statement.

We note that, Lee and Shankar~\cite{LeeShankar}
had derived a similar statement for the limited case
of hard-core models in two dimensions.
However, their argument relies on a certain field-theory mapping,
and looks less reliable compared to ours.
In fact, their statement appears to be too strong:
they state that there must be a Charge Density Wave (CDW) order
if the system with a fractional filling $\nu <1$ has a gap.
This has a rather simple counterexample:
the spontaneously dimerized ground state of a $S=1/2$
Heisenberg magnet at zero magnetization,
which corresponds to a hard-core boson system with $\nu = 1/2$,
has no long-range N\'{e}el-type (namely, CDW) order.
On the other hand, the spontaneously dimerized state
does break the translational symmetry,
and is consistent with our conclusion.

The ground state degeneracy of $S=1/2$ quantum spin systems
has been discussed in several different contexts,
for example in Ref.~\cite{Haldane2d}. 
In particular, there has been a lot of discussions on
the possibility of the exotic spin-liquid state called
the Resonating Valence Bond (RVB) state.
While various possibilities were considered under the name of RVB,
here we refer to the proposals~\cite{RVB,Sachdev}
of a disordered groundstate with a finite excitation gap,
but without any apparent breaking of the translation symmetry.
If there is really such a spin-liquid state, it appears
contradictory to our result.
This might be the reason~\cite{ALSM}
why there is no established example
of such an RVB groundstate, despite intensive search in
various  $S=1/2$ spin systems with odd number of spins per unit cell.

However, in spite of its uniform appearance,
some degeneracy is argued to exist~\cite{RVB,MSE} in the RVB state,
under the periodic boundary condition.
In Refs.~\cite{RVB,MSE}, this degeneracy is argued
to be unphysical.
For the degeneracy to be unphysical,
it must be that no physical operators distinguish
the groundstates with the spontaneously broken
translation symmetry.
While we are not sure it is possible, we do not rule out such
a possibility.
In any case, the (near-) degeneracy of the finite size system
concluded from the present argument seems consistent with
Refs.~\cite{RVB,MSE}.

The ground-state degeneracy
in a gapped phase required by the present argument,
is related to the commensurability.
Intuitively, this is quite natural;
the particles can be locked into a stable
groundstate only when it can have a commensurate structure with the lattice.
Such an intuition is, however, more or less based on some trivial
states which can be easily imagined in minds.
Nevertheless, the commensurability condition turned out to be
essential even in the presence of arbitrarily strong interaction and
quantum fluctuation,
because of the topological mechanism discussed in the present letter.

In generic cases, a finite excitation gap would be only possible
at special commensurate (rational) filling.
In case of charged particles, such a gapped phase
includes Mott and band insulators.
On the other hand, gapless phases at generic filling 
includes superfluid, Fermi liquid and possibly other
conducting phases.
In case of quantum spin systems, the gapful phases are related to the
magnetization plateau at quantized magnetization.
In any dimensions, a magnetization plateau with a finite excitation gap is
possible only if the commensurability condition is satisfied:
$n (S -m) = \mbox{integer}$ where $n$ is the number of
spins in the unit cell of the ground state,
similarly to the one-dimensional case~\cite{OYA}.
In fact, several magnetization plateaus reported in
$D>1$ dimensions (examples include~\cite{2Dplat})
satisfies this quantization condition.
However, a plateau in a magnetization curve appears
if there is no gapless excitation {\em which changes the
total magnetization}.
As already mentioned in Ref.~\cite{OYA},
the LSM argument (and the present argument) does not directly
guarantee gapless excitations of such kind.
Thus, it may be possible to have an ``exceptional''
magnetization plateau which does not obey the above simple
quantization condition,
if there are gapless excitations only with the same
total magnetization as the groundstate.
In one dimension, all plateaus should obey the quantization condition
as far as the general Abelian bosonization
treatment~\cite{OYA} is valid.
(However, see \cite{Azaria} for a possible exceptional
``plateau'' at $m=0$, and \cite{doped} for a discussion in the doped case.)
In higher dimensions, the situation is less clear,
while certainly many plateaus~\cite{2Dplat}
satisfy the simple quantization condition.
An exceptional plateau at $m=0$ might be realized in
Kagom\'{e} lattice~\cite{Kagome}, which is argued to have
singlet gapless excitations.

Finally, we note that our commensurability condition
has obvious generalizations to the spinful electron systems,
Kondo lattices, and other multi-species particle systems in
arbitrary dimensions.

\bigskip

I would like to thank Ian Affleck, Hal Tasaki and Masanori Yamanaka
for many stimulating discussions over several years.
I also thank Subir Sachdev for pointing out some relevant references.
This work is supported by a Grant-in-Aid from
Ministry of Education, Science, Sports and Culture of Japan.


\begin{references}
\bibitem{note}
While the LSM argument needs the assumption of
short-range particle hopping, there is apparently no such restriction
in the present argument.
However, an implicit assumption in Ref.~\protect\cite{Laughlin} and
in the present argument is that the finite gap does not collapse
by the AB effect.
This may be violated if very-long-range hoppings are introduced.
\end{references}
\end{document}